\documentclass{aa}
\usepackage{graphicx}
\usepackage{txfonts}
\def \etal {\textit{et al. }}

\begin{document}
    \title{Optimizing the search for transiting planets in long time series}
    \subtitle{}
    \author{Aviv Ofir \inst{1,2}}

    \institute{Institut f\"ur Astrophysik, Georg-August-Universit\"at, 
Friedrich-Hund-Platz 1, 37077 G\"ottingen, Germany. \and 
\email{avivofir@astro.physik.uni-goettingen.de}}

    \date{Received XXX; accepted YYY}


   \abstract
    {Transit surveys, both ground- and space- based, have already accumulated a large number of light curves that span several years.}
    {The search for transiting planets in these long time series is computationally intensive. We wish to optimize the search for both detection and computational efficiencies.}
    {We assume that the searched systems can be well described by Keplerian orbits. We then propagate the effects of different system parameters to the detection parameters.}
    {We show that the frequency information content of the light curve is primarily determined by the duty cycle of the transit signal, and thus the optimal frequency sampling is found to be cubic and not linear. Further optimization is achieved by considering duty-cycle dependent binning of the phased light curve. By using the (standard) BLS one is either rather insensitive to long-period planets, or less sensitive to short-period planets \emph{and} computationally slower by a significant factor of $\sim 330$ (for a 3yr long dataset). We also show how the physical system parameters, such as the host star's size and mass, directly affect transit detection. This understanding can then be used to optimize the search for every star individually.}
    {By considering Keplerian dynamics explicitly rather than implicitly one can optimally search the BLS parameter space. The presented Optimal BLS enhances the detectability of both very short and very long period planets while allowing such searches to be done with much reduced resources and time. The Matlab/Octave source code for Optimal BLS is made available\thanks{\texttt{http://www.astro.physik.uni-goettingen.de/\~{}avivofir/}}.}

\titlerunning{Optimized search for transiting planets in long time series}

\maketitle
%

\section{Introduction}

The BLS algorithm for the identification of transiting exoplanets (Kov{\'a}cs \etal 2002, hereafter KZM) is very widely used. While this work uses BLS as a concrete example, we stress that the main questions we address here are related to the information content of the data, and are therefore easily translatable to other detection algorithms. Compared to other period-searching algorithms, like Fourier transform or Lomb-Scargle periodograms, the computational load of BLS is rather high. This is an opposite conclusions to that of KZM since -- as we show here -- looking for transits requires many more frequency steps than looking for sine-like variations. Recently this situation became particularly grave since both space- and ground- based surveys now span multiple years and have obtained hundreds of thousands of light curves (if not millions) and their BLS analysis quickly becomes a challenge. In this paper we show that optimally stepping the BLS parameters, even without changing the BLS core technique, can produce very significant performance improvements \emph{and} improve sensitivity relative to standard practices at the same time.

The BLS technique fits a box of some phase duration to a light curve folded at some test frequency, at all possible starting times - producing a three-dimensional search in orbital frequency, reference phase and duty cycle. For this reason the current work is also similarly divided: after defining the duty cycle in \S \ref{DutyCycle}, we discuss the frequency search axis in \S \ref{FreqAxis}, and the reference phase and duty cycle search axes in \S \ref{Folded}. Finally, we discuss if and how to choose the best-fitting frequency in \S \ref{Choosing} and conclude in \S \ref{Discuss}.

\section{The transit signal duty cycle}
\label{DutyCycle}

BLS aims at detecting a periodic sudden and short drop in the observed flux, ostensibly caused by a transiting planet. The orbital period of a test particle, such as a planet, around the host star is given in Keplerian dynamics by:
\begin{equation}
\label{OrbPer}
 P^2= \frac{4 \pi^2}{G M} a^3
\end{equation}
Consequently the phase duration, also known as the duty cycle, of a central transit in a large ($a>>R$) circular orbit with a period $P$ (1/orbital frquency $f$) around a star with mass and radius $M$ and $R$, respectively, is :
\begin{equation}
\label{DutyCycleEq}
 q(f,M,R) \approx \frac{R}{\pi a} = \frac{(2 \pi)^{2/3}}{\pi} \frac{R}{(GM)^{1/3}} f^{2/3}
\end{equation}
Where $a$ is the orbital semimajor axis.

The observed duty cycle may be shorter than $q(f,M,R)$, for example in non-central transits, or even longer than $q(f,M,R)$ if $R$ is underestimated or $M$ is overestimated. Eccentric orbits may further change the observed duty cycle, to both shorter and longer than $q(f,M,R)$, but eccentricity cannot (usually) be detected or even constrained from photometry alone, so for the remainder of this work we assume circular orbits. Still, the quoted $q(f,M,R)$ is our a-priory expected duty cycle, and any corrections for longer or shorter duty cycles -- as discussed in \S \ref{Folded} -- should be made relative to this number.

Importantly, the typical targets for transit surveys are Solar-type stars which are all approximately similar in mass and size (a factor of 2 in either is very significant) while the orbital period can easily change over three orders of magnitudes - from sub-day to a year and longer. Therefore, for the optimization of the transit search the accuracy of the estimated $M$ and $R$ are far less important, and for the remainder of this work we assume Solar values for these parameters when giving numerical results. We note that while the original BLS did not require input of $M$ and $R$ at all, assumptions about the stellar mass and radius were always made, even if implicitly, by the definition of the other search parameters. By making them explicit we allow correcting for non-Solar values, such as in proposed searches for transiting planets around white dwarfs (Agol 2011).

\section{The frequency axis}
\label{FreqAxis}

\subsection{Search edges}
The available dataset will be described by a single parameter: its span in time, hereafter simply span and labeled $S$. The minimal search frequency is naturally derived from the goal of detecting periodic phenomena, so the maximum period can be no longer than half of the data span, or $f_{min}=\frac{2}{S}$. One may prefer to use a more conservative $f_{min}=\frac{3}{S}$ that will help to separate similar but unrelated events from true periodicity by requiring observation of a third event - but we will use the former definition below.

The maximal possible frequency could in principle be derived from the Shannon-Nyquist sampling theorem had we known the sampling frequency of our data, but a much more stringent limit usually comes from astrophysical arguments. By setting the orbital distance $a$ in eq. \ref{OrbPer} to $3R$ (scale of the Roche limit) one can determine the maximal frequency to search:

\begin{equation}
\label{OrbFreq}
f_{max}=\frac{1}{2 \pi} \sqrt{\frac{G M}{(3R)^3}}
\end{equation}
Searching the range $[f_{min},f_{max}]$ will ensure that all relevant orbital periods are covered (down to about 14.5\,hr for the Sun).

\subsection{Frequency resolution}
\label{FreqRes}
When searching for transit signals it is commonly assumed that one should uniformly sample the orbital frequency rather than the orbital period (e.g. KZM). However, while this is true for sine-like signals, including the additional information that the transit duty cycle is also frequency-dependent changes that picture. 

The frequency resolution of uniformly-sampled sine signals is $\frac{1}{S}$. In the context of experimental data this is the width of the Fourier transform peak. We wish to find the corresponding number for transit-like signals. We assume our signal was uniformly sampled and that it has an exact box shape, i.e.: when folded on the true frequency $f_{true}$ only the points in a small phase range spanning $q_{true}(f_{true},M,R)$ are all at some definite non-zero value, and all other points have a value of zero. We now consider an imperfect folding at some frequency $f_{true} + \Delta f$: in such a case each point is shifted (relative to the exact case) by $(\Delta f \cdot t)$ phase when $t$ is the data point's time. If the reference time is the beginning of the dataset then on average points move by $(\Delta f \cdot S / 2)$ phase. The probability that such a shift will move the point to outside the true box is approximately the ratio of the phase shift to the box sizes $\frac{\Delta f \cdot S/2}{q(f_{true},M,R)}$. Since the BLS score (labeled ``Signal Residue'' by KZM) scales with the number of points in transit, this is also the relative change in the BLS score due to $\Delta f$. To find the half width at half maximum (HWHM) of the BLS peak (which is a measure of frequency resolution) we set the condition for the BLS score to half its peak value: $\frac{\Delta f_{HWHM} \cdot S/2}{q(f_{true},M,R)}=\frac{1}{2}$ or: $\Delta f_{HWHM}=\frac{q(f_{true},M,R)}{S}$. In the presence of an unknown $f_{true}$ one needs to oversample this critical frequency resolution to be sure that the peak is not just missed between frequency samples. We therefore add a frequency oversampling parameter $OS$ which is the number of frequency points to calculate in the expected HWHM range to give the final 
\begin{equation}
\label{EarlyDiffEq}
  \Delta f = \frac{q(f_{true},M,R)}{S \cdot OS}
\end{equation}
$OS$ values of the order of a few (2-5) would suffice ensure the true peak is not missed. It is now easy to see that the frequency resolution $\Delta f$ is no longer constant - it depends on $f$ itself due to the physics of the problem. Also, one can compare this result to the classical frequency resolution of $\frac{1}{S}$ to find that in the framework presented here sine functions of all frequencies have a constant effective duty cycle of 1, and indeed sine waves are (almost) always non-zero.

The above result means that if one chooses a uniform frequency sampling with $\Delta f_{uniform}$ that is suitable for short-period signals (i.e., large $\Delta f$) a large portion of the long-period signals will be missed between frequency samples and will not be detected. Conversely, if one chooses a uniform frequency sampling with $\Delta f_{uniform}$ than is suitable for long-period signals (i.e., small $\Delta f$), the computation time will be artificially increased by a large factor (see next subsection).

Actually, the later choice will not just be slow to calculate but also it will be less sensitive to real signals by being too sensitive to noise: such a choice will allow ``detection'' of signals that are extremely unlikely to be transits. For e.g: a low duty cycle $q=10^{-3}$ is quite possible for a 1yr planet, but the same $q$ means a $<4.5$ minutes long transit for a 3d period - so by allowing such $q$ values one becomes sensitive to phenomena that are more likely to be noise than signals, and so the BLS background noise increases.

\subsection{Optimal frequency sampling}
Since it was shown above that a uniform frequency sampling law is not optimal, we wish to find the optimal frequency sampling sequence. This means solving the differential equation (eq. \ref{EarlyDiffEq}):
\begin{equation}
\label{DiffEq}
    \begin{array}{l}
      df=A f^{2/3} \\
      A \equiv \frac{(2 \pi)^{2/3}}{\pi} \frac{R}{(GM)^{1/3}} \frac{1}{S \cdot OS}
    \end{array}
\end{equation}

By introducing a temporary variable $x=1,2,3,...$ which is the index number of the sequence of $f$ values, we solve equation \ref{DiffEq}:
\begin{equation}
\label{DiffEqSolution}
  \begin{array}{l}
    f(x)=\left(\frac{A}{3} x + C \right)^3 \\
    C \equiv f_{min}^{1/3}-\frac{A}{3}
  \end{array}
\end{equation}

Where C is the constant of integration, and by requiring that the first value in the sequence ($x=1$) to be $f_{min}$ its value is determined. This is a non-trivial result: we have shown that for transit surveys the optimal frequency sampling is cubic in frequency, and not linear.

\begin{figure}[tbh]
   \centering
   \includegraphics[scale=0.48,bb=0 0 1200 900]{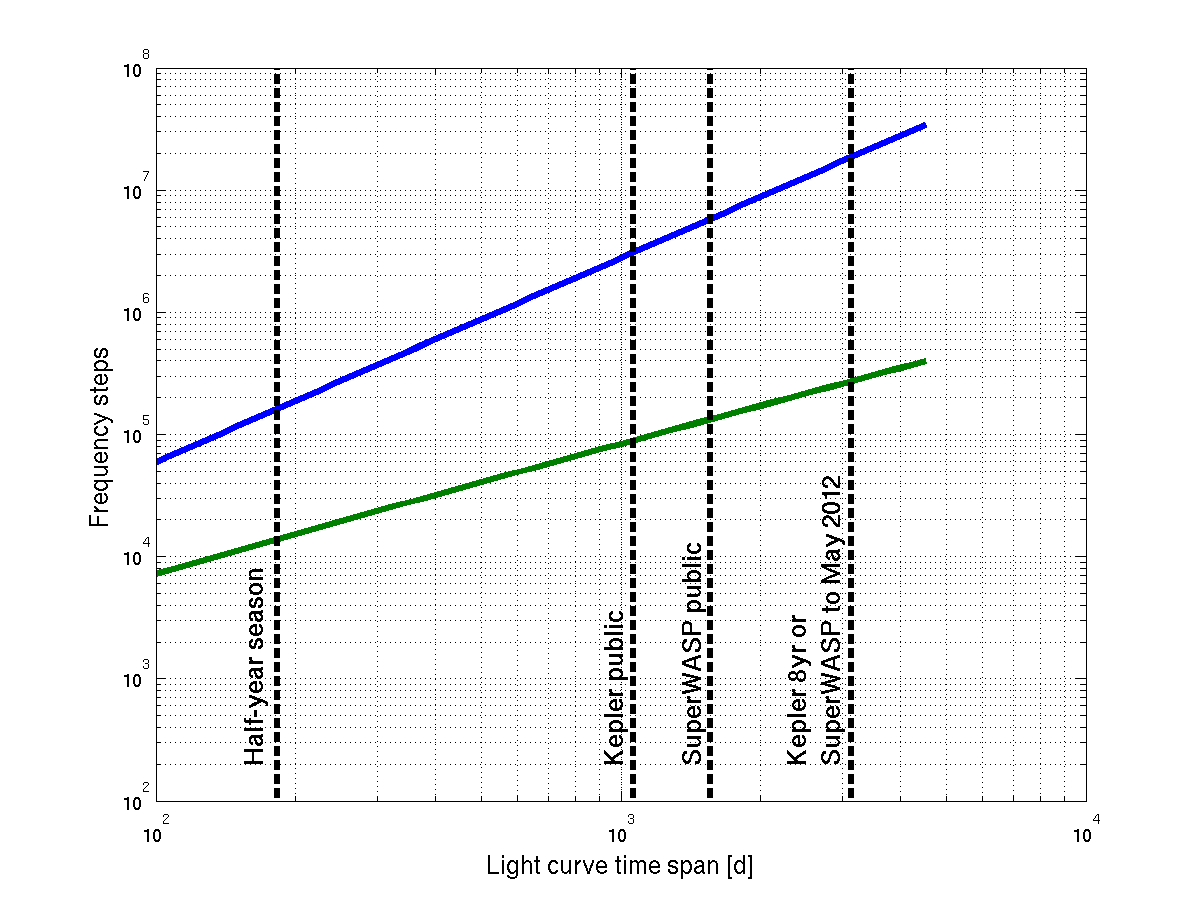}
\caption{The number of frequency steps required by a uniform sampling scheme (upper, blue line) and the optimal sampling scheme (lower, greem line). Vertical dashed lines represent the time span of some existing and expected ground- and space-based datasets.}
\label{FreqStepsFigure}
\end{figure}

Now we can count the number of frequency steps in the above optimal sampling:
\begin{equation}
\label{NfreqOpt}
 N_{freq,optimal}=\left(f_{max}^{1/3}-f_{min}^{1/3}+\frac{A}{3}\right)\frac{3}{A}
\end{equation}
For comparison, uniform sampling suitable for long periods detection has $\Delta f=\frac{q(f_{min},M,R)}{S \cdot OS}=A f_{min}^{2/3}$, which translates to number of steps of:
\begin{equation}
\label{NfreqUniform}
N_{freq,uniform}=\frac{f_{max}-f_{min}}{A f_{min}^{2/3}}
\end{equation}
For a numerical example, in the case of a 3yr light curve of a Sun-like star (which is typical of the available \emph{Kepler} data) this means that using uniform sampling is $\frac{N_{freq,uniform}}{N_{freq,optimal}}=39$ times slower than using optimal frequency sampling (see Figure \ref{FreqStepsFigure}), while not offering improved sensitivity relative to the optimal sampling (actually, the other way around, see \S \ref{FreqRes}).

Finally, we comment that since different frequencies can be evaluated independently the Optimal BLS code we make available also allows for parallel evaluation of BLS in the frequency axis on different threads \footnote{Multithreading can be controled by changing the code from  \texttt{parfor} to \texttt{for} and vice versa.}. The contribution of multithreading to the Optimal BLS performance obviously depends on the used computer system. However, we observed a less than linear performance gain when scaling the threads count, and we guess it is related to the fact that each one of the many frequencies ($>10^5$ frequencies for a 3yr data) takes only a very short time to calculate so the communication overhead is significant. We therefore recommend using multithreading only for the the manual evaluation of one/few objects, which is required when more attention is given to specific systems. For large scale processing we recommend using the different threads to process different groups (which requires no communication, and also strengthens the case for not-too-large groups, where the concept of BLS groups is explained in \S \ref{PhaseFolding}).

\section{On the folded light curve}
\label{Folded}

KZM showed that computing the BLS metric using the original data points is computationally wasteful since virtually the same result can be obtained using phase-binned data -- as long as the bin size is small enough so that at least one bin would fall completely in-transit, and not only partially so. We therefore describe below the binless and binned BLS parameters, where the former describes more accurately the possible true signals and the later is the corresponding parametrization of Optimal BLS.

\subsection{Binless BLS:} 

\textbf{Duty cycle axis:}  In this section we assume that the data was already folded at some test frequency $f_{test}$. We therefore expect that a central transit would have a duty cycle of $q(f_{test},M,R)$. However, as mentioned above, the observed transit can be both longer or shorter. Importantly, extreme circumstances are required for the observed transit to be more than a few times longer or shorter than the expected $q(f_{test},M,R)$. We can therefore parameterize factors  $1/Q_{min}>1$ and $Q_{max}>1$ such that we wish to look at test duty cycles $q_{test}$ in the range
\begin{equation}
 \frac{1}{Q_{min}} < \frac{q_{test}}{q(f_{test},M,R)} < Q_{max}
\end{equation}
and expect that both $1/Q_{min}$ and $Q_{max}$ to have a value of a few: probably no more than about 3, and with only extreme cases reaching to 5 or higher values.

\textbf{The reference phase axis:} the reference phase (the mid-transit phase) can have any value between 0 and 1, and all points should be checked in principle. However, there is little gain in setting the reference phase resolution to much smaller then $q(f_{test},M,R)$ itself, since the number of in-transit points will not change significantly.

\subsection{BLS with binning:} 

\textbf{Duty cycle axis:} When one bins the data one loses some sensitivity by not optimally using the data in the bins that contain either the ingress or the egress. This effect can be minimized by using a large number of small bins, but then the computational benefit of binning erodes. We show how to properly choose the bin size to find a good balance between the speed of computation and sensitivity. The binless $Q_{min}$ and $Q_{max}$ parameters above are used to define their binned counterparts $MinBin$ and $NumBins$ in the following way: since one needs at least one bin to be completely in-transit the minimum bin size should be smaller than the smallest expect transit $q(f_{test},M,R)/Q_{min}$ by some factor $MinBin$. Larger expected transits up to $Q_{max}$ would be fitted using an integer number of this minimum-sized bin given by $NumBins$. Mathematically this means that

\begin{equation}
  \textrm{smallest bin size}=\frac{q(f_{test},M,R)/Q_{min}}{MinBin}
\end{equation}
$MinBin$ should be larger than 2 to be sensitive to transits $q(f_{test},M,R)/Q_{min}$ long regardless of the true transit phase, and should have a value of a few (i.e., not a large number) to reduce the computational load. To avoid confusion, we stress that $Q_{min}$ and $Q_{max}$ describe the underlying transit signals that we want to search for, while $MinBin$ and $NumBins$ describe the search grid.

Longer transits will be searched using groups of bins, or a bin width of an integer multiple of this smallest bin size. The longest expected transit would have a duration that is $Q_{min} \cdot Q_{max}$ times the shortest expected transit, so the largest number of bins to be searched is
\begin{equation}
\label{NumBinsEq}
  NumBins=Q_{min} \cdot Q_{max}
\end{equation}

\textbf{The reference phase axis:} Using such a binning scheme would allow signals to be readily detected even if their reference time is shifted with respect to the used bins grid, since there will always be at least one bin which is completely in-transit. One therefore still has to check all possible reference phases, but now at a reduced resolution of the bins, which is automatically not much smaller than $q(f_{test},M,R)$.

\subsection{The computational load}
The configurations that need to be checked are all $NumBins$ possible widths, starting from all (smallest\,bin\,size)$^{-1}$ possible reference phases. Importantly, when optimally binning, the computational load of the individual test frequencies also depends on $q(f_{test},M,R)$, meaning that again the choice of a constant bin size incurs penalties. Specifically, the number of configurations to check by using optimal binning is:
\begin{equation}
  N_{conf., optimal}(f_{test})=NumBins \cdot \frac{MinBin \cdot Q_{min}}{q(f_{test},M,R)}
\end{equation}
A uniform binning scheme suitable for lone-periods detection would have to scan a different number of configurations:
\begin{equation}
  N_{conf., uniform}(f_{test})=NumBins \cdot \frac{MinBin \cdot Q_{min}}{q(f_{min},M,R)}
\end{equation}
Note that by definition of this scheme as uniform it does not have dependency on $f_{test}$.
The total computational load is just the integration of the number of configurations that need to be checked over all frequencies:
\begin{equation}
  N_{conf., optimal}=\frac{NumBins \, MinBin \, Q_{min}}{A^2 \, S \, OS} \cdot 3\left(f_{min}^{-1/3}-f_{max}^{-1/3}\right)
\end{equation}
\begin{equation}
  N_{conf., uniform}=\frac{NumBins \, MinBin \, Q_{min}}{A^2 \, S \, OS} \cdot \frac{f_{max}-f_{min}}{f_{min}^{4/3}}
\end{equation}
The total speedup of using optimal sampling and binning is therefore:
\begin{equation}
\label{speedup}
  \textrm{speedup}=\frac{N_{conf., uniform}}{N_{conf., optimal}}=\frac{{f_{max}-f_{min}}}{3f_{min}^{4/3}\,\left(f_{min}^{-1/3}-f_{max}^{-1/3}\right)}
\end{equation}
Which for a 3yr dataset is a very significant speedup factor of 337. For a 8yr dataset the speedup factor is even higher at $>880$.

\subsection{Phase folding}
\label{PhaseFolding}

Apart from discussing the BLS calculation on the folded light curve we note that the phase-folding step itself is computationally intensive. When applied to simulated continuous 3yr long 
\emph{Kepler}-like dataset, our optimized implementation spent $\sim 2/3$ of the time phase-folding the data. Since for the vast majority of targets in a transit survey there is near-perfect overlap in time stamps (every image of a given field is common to nearly all its targets), these are mostly redundant calculations. Importantly, these redundant calculation will significantly dilute the benefits of Optimal BLS and cause the actual run-time to be longer than suggested by eq. (\ref{speedup}) since phase-folding of the data is sped up only by the factor given by dividing eq. \ref{NfreqUniform} by eq. \ref{NfreqOpt}.

We therefore allowed for parallel execution of all the BLS calculation steps on multiple stars -- so multiple stars are evaluated at each test frequency -- and the phase folding is done once for each such group. From practical standpoint, such a group should typically contain several tens of stars (maybe up to a hundred or so) which virtually nullifies the relative time penalty for phase folding, but without incurring a memory penalty by including too large groups. Avoiding too large groups is advisable also since one may not always find enough stars with similar-enough masses and radii (which affect the frequency sampling and phase binning schemes).

\begin{figure}[htb]
   \centering
   \includegraphics[scale=0.4,bb=0 0 814 641]{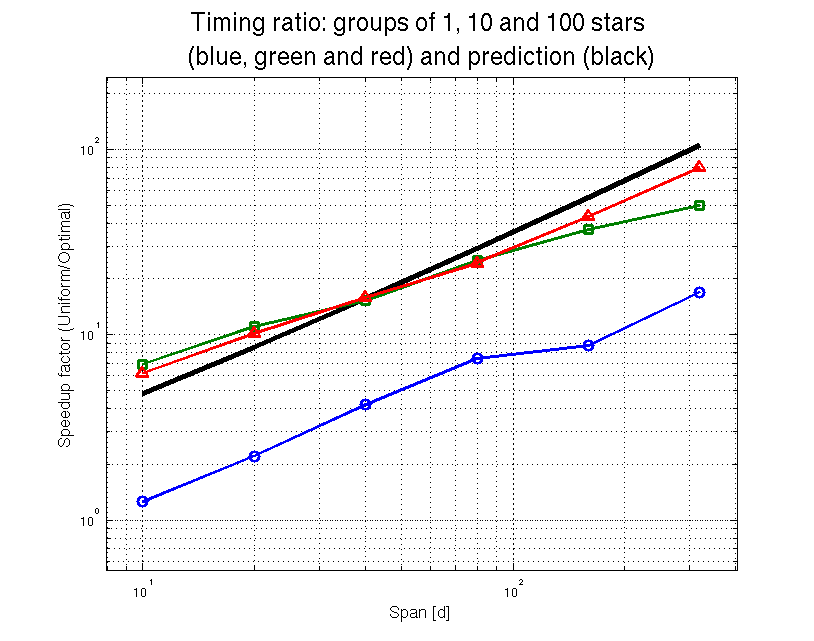}
\caption{The speedup factor of Optimal BLS relative to the un-optimized BLS vs. the span of the simulated dataset (10, 20, 40, 80, 160 and 320 days). The different lines are the ratio of run times measured for different runs, each one for a different group size (1, 10 and 100 stars per group, in blue circles, green squares and red triangles respectively). The solid black line is the predicted speedup factor from e.q. \ref{speedup} \emph{and is not a fit}. See text for further discussion.}
\label{SpeedupFigure}
\end{figure}

When comparing the actual time it takes to calculate BLS on one group using our original and optimized versions, one would expect that due to the above effects the observed speedup factor will be lower than that predicted by eq. \ref{speedup} for ``groups'' of one light curve. On the other hand, the observed speedup factor will approach eq. \ref{speedup} for large-enough groups since for them the relative time spent on phase-folding was reduced. This expectation, as well as the speedup scaling with data span (which is the main result of this work), were observed in actual calculations spanning a factor $>30$ in data spans and a factor of 100 in group sizes (Figure. \ref{SpeedupFigure}), validating the above analysis.

\section{Choosing the best-fitting frequency}
\label{Choosing}

\subsection{Normalization}

Obviously, the best-fitting frequency is the one with the highest significance above the background of other tested frequencies. We now consider how to choose this best-fitting frequency, and determine it's significant. The first question appears to be trivial as one may simply answer that it would be the frequency with the corresponding highest numerical value of the above BLS run. A first-order answer to the second question was given already by KZM: they defined the SDE (Signal Detection Efficiency) metric by first subtracting the mean of the BLS values and than normalizing to the scatter of the resultant data. This made the SDE naturally scaled: an SDE value of X means that the this particular period is more significant by $X \sigma$ than the bulk of checked orbital frequencies.

However, the BLS spectrum of real targets has two important structures that make these choices sub-optimal: (1) there is a rising trend of the BLS values towards lower frequencies, i.e.: the local average of the periodogram rises. (2) there is a rising scatter towards lower frequencies, i.e.: the local scatter around the local average of periodogram increases towards lower frequencies. These two frequency dependencies are expected: as datasets get longer the tested orbital frequencies get smaller and so the expected duty cycle at each of these test frequencies $q(f_{test},M,R)$ gets smaller. This means that a progressively smaller number of points are actually required to be in-transit, and so the probability that the mean value of smaller random groups of points will be different from the global mean will increase, driving both the mean value and the scatter of the raw BLS spectrum up. Also, the sensitivity to a small number of anomalies or ill-corrected discontinuities will be significantly increased as test frequencies decrease as there will be less data points ``in-transit'' to average-out such instances. All the above will result in an accelerating rate of changes in the ``background'' BLS spectrum towards lower frequencies.

In order to solve these problems we slightly generalize the way SDE is calculated:
\begin{itemize}
 \item Instead of removing the mean value, we remove the median-filtered periodogram. We note that due to the optimal sampling of the frequency axis, peaks at all frequencies now include a similar number of points, so by choosing a median filter with window size much larger ($\ge 10$ times) than the oversampling parameter $OS$ one ensures the peaks of interest are not filtered or attenuated. We label this curve as $\textrm{BLS}_{ternd}(f)$.
 \item After removing $\textrm{BLS}_{ternd}$ we wish to evaluate the local scatter of the data. For that we simply evaluate $abs(diff(\textrm{BLS}))$ -- the absolute value of the point-to-point difference of the periodogram. We then median-smooth that data similarly to the way it was done above, to obtain the local scatter curve - and we label this as curve as $\textrm{BLS}_{scatter}(f)$.
 \item Our genaralized SDE is now: $\textrm{SDE}(f)=\frac{\textrm{BLS}(f)-\textrm{BLS}_{ternd}(f)}{\textrm{BLS}_{scatter}(f)}$
\end{itemize}

\begin{figure*}[htb]
   \centering
   \includegraphics[scale=0.43,bb=0 0 1486 553]{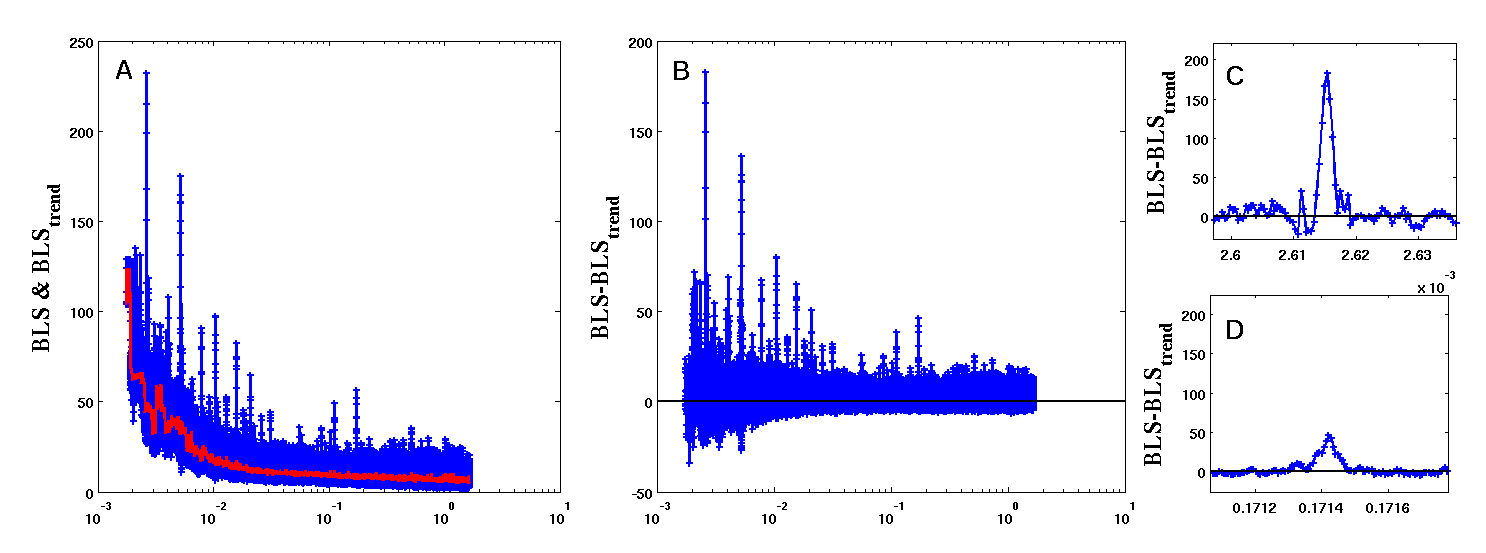}
\caption{Panel A: The Optimal BLS spectrum of \emph{Kepler}'s KOI 1574 (blue) and its $\textrm{BLS}_{ternd}$ (red) after removing planet candidate KOI 1574.01 using all available data through Quarter 13. Panel B: The same, after removing $\textrm{BLS}_{ternd}$. Panels C, D: two zoomed-in regions with significant local peaks. Optimal frequency sampling is evident by the fact that each peak is sampled with similar number of samples despite the large difference in width.}
\label{RawBLS}
\end{figure*}

Now the SDE is normalized to better reflect the true significance of a given peak over its local background, even if one chooses a different technique of estimating the local BLS trend and BLS scatter. Some of these effects can be see on Figure \ref{RawBLS}: this is the raw Optimal BLS spectrum of a typical target (\emph{Kepler}'s KIC 10028792, or KOI 1574) spanning 1141d, after SARS de-correlation (Ofir \etal 2010) and removal of its known giant planet candidate (KOI 1574.01, Batalha \etal 2012). It is easy to see that:
\begin{itemize}
  \item There is a sharp rise of BLS towards low frequencies: the BLS trend is near-constant above $10^{-2} d^{-1}$, or on more than $99\%$ of the frequency range, and practically all the BLS trend variability is below that threshold. 
  \item The local scatter is also increasing towards lower frequencies -- from a value of about 1.2 to a value of 3 and more in this particular case. This is best seen by looking at the local scatter of the BLS trend curve.
  \item The highest peak (at $f=2.615\cdot 10^{-3} d^{-1}$ is the first harmonic of of the 191d planet candidate reported by Ofir and Dreizler (2012). Notably, the width of the peak is a few times $10^{-6} d^{-1}$ (panel C). This width can be compared to another peak (panel D) at $P \sim 5.83d$ ($f\sim 0.171d d^{-1}$) which has a width of $\sim 10^{-4} d^{-1}$, or several tens of times wider. With SDE$>32$ this additional peak is not noise: it (and the adjacent peak at $f\approx 0.111$ too) correspond to two new candidate that were not detected by the Kepler pipeline\footnote{At the time this paper was submitted} and are now reported in [Ofir \etal 2013], despite their raw BLS score being much lower than the corresponding low-frequency background. 
  \item Apart from the general trend, there are several additional rather sharp breaks in the BLS spectrum (especially visible on the trend curve). Due to these jumps the BLS trend curve cannot be well-modeled by a simple function like a polynomial, and a numerical function like a median filter (but not limited to it) may perform better.
\end{itemize}

The last remaining issue is to determine the threshold value above which candidate signals should be considered as probable real detections.

\subsection{Significance threshold}
\begin{figure}[htb]
   \centering
   \includegraphics[scale=0.46,bb=0 0 720 543]{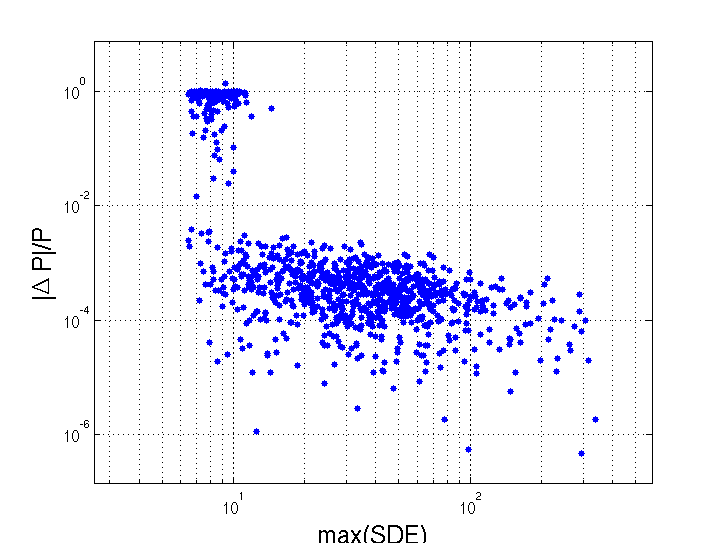}
\caption{The (absolute value) fractional difference between injected and detected period, vs. SDE for 1000 simulated signals in 100d long light curves. The vertical separation between good detections (lower points cloud) and misses (upper points cloud) increases as the data span increases, and here these are already well separated. Importantly, there is not a single missed signal with SDE$>15$.}
\label{SDE}
\end{figure}

We simulated 1000 light curves, all spanning 100d and containing transit signals of various periods, durations and depths both above and below the detection threshold. The light curves were rather benign in the sense that they included white noise only, and no jumps or discontinuities were simulated. We then compared the detected and simulated periods, and plotted the absolute value of their fractional difference against the SDE (as defined above) of that detected period (see fig. \ref{SDE}). The separation between good detection and misses is clear. Importantly, all the misses also had low SDE, so the false alarm rate (a detection when there is no signal) is very low for SDE$>15$. We conclude that detections of approximately this significance and higher should be promoted to the next level of checks (which is beyond the scope of this paper).

\begin{table*}
\caption{Parameter definition and source of value.}
\begin{tabular}{l l l}
\hline
Parameter name	& Meaning			& Value choser by \\
\hline
$M$ [kg]	&  target star mass 		& external data (if available) \\
$R$ [m]		&  target star radius 		& external data (if available) \\
\hline
$S$ [s]		&  span of data			& From the light curve \\
$f$ [s$^{-1}$]	&  test frequency		& eq. \ref{DiffEqSolution} \\
$f_{min}$ [s$^{-1}$] & lowest searched frequency	& 2/$S$ or 3/$S$\\
$f_{max}$ [s$^{-1}$] & highest searched frequency	& eq. \ref{OrbFreq} \\
$OS$ 		&  frequency oversampling	& order of a few (3 - 5) \\
$Q_{min}$	&  shortest expected transit relative to $q$	& order of a few (3 - 5) \\
$Q_{max}$	&  longest expected transit relative to $q$	& order of a few (3 - 5) \\
$MinBin$	&  smallest used bin size relative to $Q_{min}$	& order of a few (3 - 5) \\
$NumBins$	&  maximum number of smallest bins & eq. \ref{NumBinsEq} \\
\hline
\end{tabular}
\label{ParamsTable}
\end{table*}

\section{Discussion}
\label{Discuss}

The \emph{Kepler} public archive \footnote{http://archive.stsci.edu/kepler/} currently includes about 200,000 stars monitored for about three years. The ground-based SuperWASP survey public archive \footnote{http://www.wasp.le.ac.uk/public/} is much larger, holding almost $1.8\cdot10^7$ unique light curves and spanning a longer time base (though not continuously as \emph{Kepler}), and there are many other publicly available datasets. For scale, we note that the \emph{Kepler} transiting planets search is performed on the Pleiades -- NASA's most powerful supercomputer -- which runs at over one petaflop
. These resources are hardly commonly available, so if one wishes to make an independent search for transiting planets in the \emph{Kepler} dataset (e.g. Ofir \& Dreizler 2012) one should try to use the available resources efficiently. In this paper we show that by assuming Keplerian dynamics one can simultaneously reduce the computation time by two to (almost) three orders of magnitude and improve the sensitivity for shallow transit signals.

We explained in detail the physical meaning of all parameters to aid the understanding of the parameters used in the context of the problem of transit detection. The range of freedom for the algorithm to try fitting many box-shaped models to the data is given in the form of a series of normalized parameters that all have a suggested value of a few. This formulation allows for a ``hands-free'' application of BLS - it will be applied optimally fast and with optimal sensitivity to all light curves regardless of their span. Furthermore, if one happens to have external information on the target stars (roughly estimated mass and/or radius) the information can be incorporated to further tailor the search for this particular target, including non-standard configurations such as planets around white dwarfs (Agol 2011). Importantly, the stellar parameters are not new - they were simply implicitly assumed when setting the search parameters on the unoptimized BLS.

Importantly, significant speedup can be obtained by using unmodified BLS code by simply sampling the frequency axis optimally (see Figure \ref{FreqStepsFigure}) instead of uniformly. Further optimizations can be attained by optimal binning (small changes to the code) and common phase folding (larger modifications). In any case, we make the Matlab/Octave source code for Optimal BLS freely available. Another result of formulating all parameters relative to Keplerian orbits is that well-understood trade-offs are now possible. For example, one may be willing to lose some sensitivity to planets with long transits caused by high eccentricities -- these are likely to be rejected later as suspected EBs -- in order to speed up calculations.

While the importance of optimizing the search for long-period planets is almost obvious, we note that the improved sensitivity to short-period signals recently became also important as a new sub-population of very hot and very small planets -- that do not exist on the Solar system -- is now being detected. The first members in this group were CoRoT-7b and Kepler-10b (L{\'e}ger \etal 2009 and Batalha \etal 2011, respectively), but now it also includes a number of compact multi-planet systems that all have periods of a few days or even less than a day, such as the KOI 961 system (Muirhead et al.2012).

\section{Acknowledgements}

I acknowledge financial support from the Deutsche Forschungsgemeinschaft under DFG GRK 1351/2. I thank Guillem Anglada and Mathias Zechmeister for fruitful discussion of these results. The starting point of this work is a MATLAB implantation of BLS that was developed by Omer Tamuz and Tsevi Mazeh.

\end{document}